# On a New Method of Storing a Variable Size Array

Anatolijs Gorbunovs
University of Latvia
Riga, 2012

# Abstract

There are several known ways of storing stack data structure – as a linked list (called list from now on in this paper), an array which size is dynamically increased on demand (called vector), or a double-ended queue (called deque).

C++ Standard [1] suggests a deque as a standard container to use as a back-end for a stack's internal storage. C++ deque implementations should rely on memory paging for a better use of processor's caching mechanism [2] (though on practice this advice is not always followed).

This paper introduces a new data structure, log_vector, with the following properties:

- constant time random access to individual elements;
- constant time element addition to the end;
- constant time element removal from the end;
- constant time empty data structure creation;
- amortized constant space per individual elements;
- constant additional space used;

In theory it outperforms vector in the element addition and removal, behaves similarly to deque on all comparison parameters listed here, and has better random access time than list.

It will be shown that on practice it behaves better than list, deque and vector when used as a stack with indexed access to its elements.

## Problem

The main problem of vector is that it keeps copying elements when it goes out of allocated memory. Thus, an element addition may require as much as O(N) time, where N is current size of the vector [3].

Problem of the deque, on the other hand, is that it allocates memory in a chunks of constant size, which are usually too small (because if they were too big, the space complexity would not be constant per element) [2].

Linked list has poor caching possibilities, because memory is allocated in different places. It also suffers from too often memory allocation [4].

## Idea

The idea is to allocate memory in chunks of size equal to powers of 2, like vector does, but when a memory becomes too low to store a new item and we have to increase the internal storage, not to copy-and delete the previously allocated chunk, but just allocate another chunk of size equal to sizes of all the chunks previously allocated, plus 1.

This model allows relatively easy access to individual elements by their indexes, fast increasing of internal storage (just allocation, no copying) and decreasing of internal storage (just deallocation, no copying). This is all as good as deque does. Comparing to deque, it solves the problem of too small/large chunk sizes.

## Storage

The data structure has 4 fields:

- size – the total amount of elements currently stored in the data structure;
- ptrs – an array of $\log_2$(MAX_SIZE) pointers to chunks;
- chunks – number of already allocated chunks;
- active_chunks – number of active chunks: either equals to chunks, or to chunks - 1.

A picture illustrates how this works:

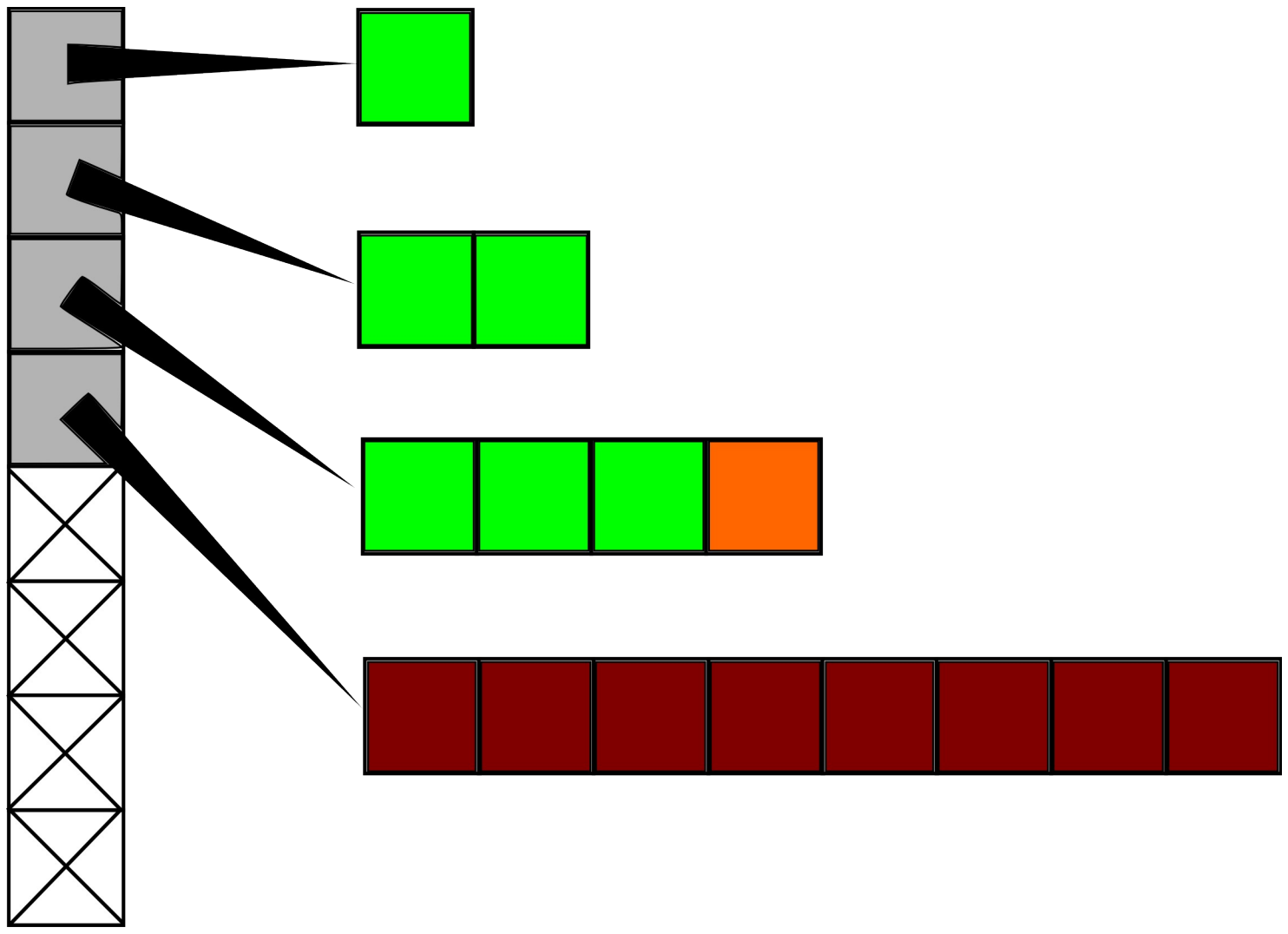

Here, at the left size of the picture are 8 pointers of ptrs array, 4 bottommost of them are null pointers. The other 4 topmost point to the so-called chunks of sizes equal to increasing powers of 2 (1, 2, 4, 8). Their amount equals to chunks variable.

Green squares store the actual data. They belong to so-called active chunks. Amount of chunks that have at least one green square equals to active_chunks.

Red squares belong to a deactivated chunk – previously it was added, because there was not enough storage for the elements (i. e. there was at least one green square in it at some point in time).

Total amount of green squares equals to size variable.

## Indexes

To determine, an number of a chunk *c*, to which an element at global (external) index *i* belongs, we use the following formula.

$c$ = MSB($i$ + 1)

Here, we use a MSB operation, which returns the most significant bit of a number. It is available in most modern computer instruction sets.

To calculate an index of the element inside the chunk, let us denote this index *k*, we use the following formula.

$k = (i + 1) \oplus 2^c$

Why does this work?

Le us first examine the first formula. The total number of all the elements in the chunks before the wanted one is $2^0+2^1+2^2+...+2^{\lceil \log_2 i \rceil -1}$ , that is, sum of powers of 2, each of that is less than or equal to *i*. It is known that such a sum is equal to the $2^{\lceil \log_2 i \rceil}-1$ . To prove the formula, we will go from the opposite direction.

With the MSB($i$ + 1) there are 2 cases: either i in binary is all ones, or not. The later is a common case; it follows that adding one does not change the MSB result. Thus, we have c equals to the maximal power of 2, which is less than the index *i*. It follows from the definition of our data structure that the total amount of elements in chunks before *c*-th will be $2^{\lceil \log_2 i \rceil}-1$ . We notice that this is essentially equal to the number of elements in chunks before the wanted chunk, thus the chunk numbered *c* is the wanted one. The second case is different: adding 1 to the binary representation of *i* will definitely change the MSB value. This is good, because it may be seen from the picture and proven by analogy with a previous case that the chunk number should be increased by one indeed, comparing to the first case' formula.

The formula for index of element inside the chunk is proven similarly. It is an exercise to the reader. I may put the proof later in here, though.

## Popping

After removing the last element of the list, a situation where the last layer is empty, may occur. This case is ignored for some time, and the last layer is just marked as deactivated. It should be noted that this last non-active layer may be one (or none at all). To accomplish this, when there is one more level going to be deactivated, the currently deactivated layer should be deleted completely (deallocated), and the just emptied level should be marked as inactive.

This provides a possibility to process a situations, when there push and pop operations are alterating.

Practically this means that up to 75% of the data structures' allocated memory will be wasted. This is relatively good number, taking into account the rival data structures' parameters.

## Pushing

Pushing of the element into the end of the list should be done in such a way that deactivated level invariant is kept. To be more specific, first, we try to use (and put element into) the existing active level (which is partially filled); if the active layer if full, then we try to reuse the previously deactivated level – activate it and fill its zeroth element; allocation of a new level occurs only if there is no deactivated level available, in this case the element is pushed into the zeroth slot in this new level.

## Construction

A construction of a non-empty data structure and its filling with default values can be optimized by a constant factor.

The idea is to create a layers one-by-one and fill them in a for loop of with a platform-specific memory filling function (if available). Amount of elements yet to initialize should be kept and decreased by the amount of elements in a given layer. The initialization is done when the remaining size is zero or negative.

## Iterating

In order to iterate over the data structure, the layer number, element number in the layer as well as a reference to the data structure or its element table should be kept.

In order to iterate to the next element, we increase the element order number in the layer. If it matches total element count in the layer, we increase the layer counter and set the element counter to zero.

Reverse iterating goes analogously – by subtracting one from the element order number in the layer. It -1 is received, that means that we should go to the previous layer, setting the element number to the element total count in it minus one.

## Technical details

The total amount of elements in a layer is $2^{level\ order\ number}$. C and C++ languages allow to write it as (1 << *c*), where *c* – level number (starting from 0-th).

MSB (Most Significant Bit) operation is not available in the GNU GCC environment. The __builtin_clz can be used instead, it returns an amount of leading zeroes in an integer. Correspondingly, subtracting this amount from the count of bits in the number gives MSB value.

MS Visual Studio has _BitScanReverse function, which stores MSB in one of its parameters.

## Implementation

A reference implementation in C++ follows.

```
/*
 * File:   log_vector.h
 * Author: Anatoly Gorbunov
 *
 * Created on September 13, 2011, 9:44 PM
 */

#ifndef LOG_VECTOR_H
#define         LOG_VECTOR_H

#include "algorithm"

template<typename T>
class log_vector {
private:
        static const int MAX_PTRS = 31;
        T *m_ptrs[MAX_PTRS];
        unsigned m_size;
        unsigned m_chunks;
        unsigned m_active_chunks;
        // Size = 1, 2, 4, 8, 16, 32, ...
        // Sums = 1, 3, 7, 15, 31, 63, ...

        // Invariant:
        // Chunks in range [0, m_active_chunks - 1) are fully filled.
        // Chunk (m_active_chunks - 1) is partially filled or also fully filled.
        // m_chunks is the amount of allocated arrays.
        // m_chunks == m_active_chunks || m_chunks == m_active_chunks + 1.

        static unsigned most_significant_bit(unsigned x) {
#ifdef _MSC_VER
                unsigned long result = 0;
                _BitScanReverse(&result, x);
                return (unsigned)result;
#else
                return (sizeof(unsigned) << 3) - 1 - __builtin_clz(x);
#endif
        }

        static void calculate_iterator(unsigned global_index, unsigned &chunk,
                unsigned &index)
        {
                // Increase global_index only once.
                global_index++;
                chunk = most_significant_bit(global_index);
                index = (global_index ^ (1 << chunk));
        }

public:
        class iterator {
        private:
                T **m_ptrs;
                unsigned m_chunk;
                unsigned m_index;

                // Invariant:
```

```
		// global_index == (1 << m_chunk) - 1 + m_index
		// In binary:
		// global_index == 000000000101110010101110100111O
		//                ^
		//         most_significant_bit == 21
		// m_chunk == most_significant_bit(global_index)
		// m_index == global_index ^ (1 << m_chunk)
		// There is one special case, though:
		// When an index == 0, and global_index == (all ones in binary),
		// then a chunk is by 1 greater than when calculated by this formula.

		iterator(T **ptrs, int chunk, int index) :
			m_ptrs(ptrs), m_chunk(chunk), m_index(index)
		{ }

		iterator(const iterator &it) :
			m_ptrs(it.m_ptrs), m_chunk(it.m_chunk), m_index(it.m_index)
		{ }

		friend class log_vector;

	public:
		iterator operator++() {
			iterator it(*this);
			if (++m_index == (1 << m_chunk)) {
				m_chunk++;
				m_index = 0;
			}
			return it;
		}
		iterator operator++(int) {
			if (++m_index == (1 << m_chunk)) {
				m_chunk++;
				m_index = 0;
			}
			return *this;
		}
		iterator operator--() {
			iterator it(*this);
			if (!m_index--) {
				m_chunk--;
				m_index = (1 << m_chunk) - 1;
			}
			return it;
		}
		iterator operator--(int) {
			if (!m_index--) {
				m_chunk--;
				m_index = (1 << m_chunk) - 1;
			}
			return *this;
		}
		iterator operator +=(int shift) {
			calculate_iterator(
				(1 << m_chunk) - 1 + m_index + shift, m_chunk, m_index);
			return *this;
		}
		iterator operator -=(int shift) {
			return operator +=(-shift);
		}
		iterator operator +(int shift) const {
```

```
            return iterator(*this) += shift;
        }
        iterator operator -(int shift) const {
            return iterator(*this) -= shift;
        }
        T &operator *() {
            return m_ptrs[m_chunk][m_index];
        }
        const T &operator *() const {
            return m_ptrs[m_chunk][m_index];
        }
    };

    log_vector(unsigned size = 0, const T &default_value = T()) :
        m_size(size), m_chunks(1)
    {
        // Allocate the first chunk.
        m_ptrs[0] = new T[1];

        // If there are 0 elements, the first chunk will be inactive.
        m_active_chunks = size > 0;

        // Remove the first chunk's single cell from the size.
        // Size may become -1 at this point, which becomes unsigned 0XFFFFFFFF.
        if (size) {
            m_ptrs[0][0] = default_value;

            int s = size - 1;

            // While size is bigger than total amount of cells in all the
            // chunks, keep decreasing it (size) by an amount of cells in
            // the current chunk.
            while (s > 0) {
                s -= 1 << m_chunks;
                m_ptrs[m_chunks] = new T[1 << m_chunks];
                std::fill(m_ptrs[m_chunks], m_ptrs[m_chunks] + (1 << m_chunks),
                    default_value);
                m_chunks++;
                m_active_chunks++;
            }
        }
    }

    ~log_vector() {
        clear();
        // Do not forget to clear the last chunk.
        delete[] m_ptrs[0];
    }

    void clear() {
        // Leave the first chunk.
        while (m_chunks > 1) {
            delete[] m_ptrs[m_chunks - 1];
            m_chunks--;
        }
        m_active_chunks = 0;
        m_size = 0;
    }

    unsigned size() const {
        return m_size;
```

```
}

T &operator[](unsigned index) {
	unsigned chunk, i;
	calculate_iterator(index, chunk, i);
	return m_ptrs[chunk][i];
}
const T &operator[](unsigned index) const {
	unsigned chunk, i;
	calculate_iterator(index, chunk, i);
	return m_ptrs[chunk][i];
}

void push_back(const T &x) {
	// If have to add another layer.
	if ((1 << m_active_chunks) - 1 == m_size) {
		// If there are no previously disabled layers.
		if (m_chunks == m_active_chunks) {
			m_ptrs[m_chunks] = new T[1 << m_chunks];
			m_chunks++;
		}
		// Else, we already have a disabled layer, which should be
		// activated.
		m_active_chunks++;

		// Assign a value here to save some index calculations.
		m_ptrs[m_active_chunks - 1][0] = x;
	}
	// Do not need to add a layer, have enough space in the last active
	// layer.
	else {
		m_ptrs[m_active_chunks - 1]
			[m_size - ((1 << (m_active_chunks - 1)) - 1)] = x;
	}
	m_size++;
}
void pop_back() {
	// May have to remove not the topmost layer, but the previously
	// deactivated layer that is one level higher than the topmost,
	// and deactivate the topmost one.
	if (1 << (m_active_chunks - 1) == m_size) {
		// If already have a deactivated layer, delete it.
		if (m_active_chunks == m_chunks - 1) {
			delete[] m_ptrs[m_chunks - 1];
			m_chunks--;
		}
		// Deactivate currently active layer.
		m_active_chunks--;
	}
	m_size--;
}

T &front() {
	return m_ptrs[0][0];
}
const T &front() const {
	return m_ptrs[0][0];
}
T &back() {
	return m_ptrs[m_active_chunks - 1]
		[m_size - 1 - ((1 << (m_active_chunks - 1)) - 1)];
```

```
	}
	const T &back() const {
		return m_ptrs[m_active_chunks - 1]
			[m_size - 1 - ((1 << (m_active_chunks - 1)) - 1)];
	}

	iterator begin() {
		return iterator(m_ptrs, 0, 0);
	}
	iterator end() {
		return iterator(m_ptrs, m_active_chunks - 1,
			m_size - 1 - ((1 << (m_active_chunks - 1)) - 1));
	}
};

#endif /* LOG_VECTOR_H */
```

# Performance

First, a small data sizes – int structures were tested, using various data structures as a stack. Results are shown in the table and diagrams below.

| log_vector, MS V | log_vector, MinGW | vector, MS VC++ | vector, MinGW G- | deque, MS VC++ | deque, MinGW G- |
|---|---|---|---|---|---|
| 0.000002020 | 0.000004830 | 0.000004050 | 0.000005930 | 0.000007650 | 0.000001870 |
| 0.000007800 | 0.000012500 | 0.000012500 | 0.000010900 | 0.000079500 | 0.000012500 |
| 0.000063000 | 0.000125000 | 0.000156000 | 0.000109000 | 0.000749000 | 0.000093000 |
| 0.001250000 | 0.000780000 | 0.002030000 | 0.001880000 | 0.007950000 | 0.001090000 |
| 0.014000000 | 0.012500000 | 0.024900000 | 0.017100000 | 0.092100000 | 0.012500000 |
| 0.140000000 | 0.124000000 | 0.266000000 | 0.203000000 | 0.951000000 | 0.156000000 |

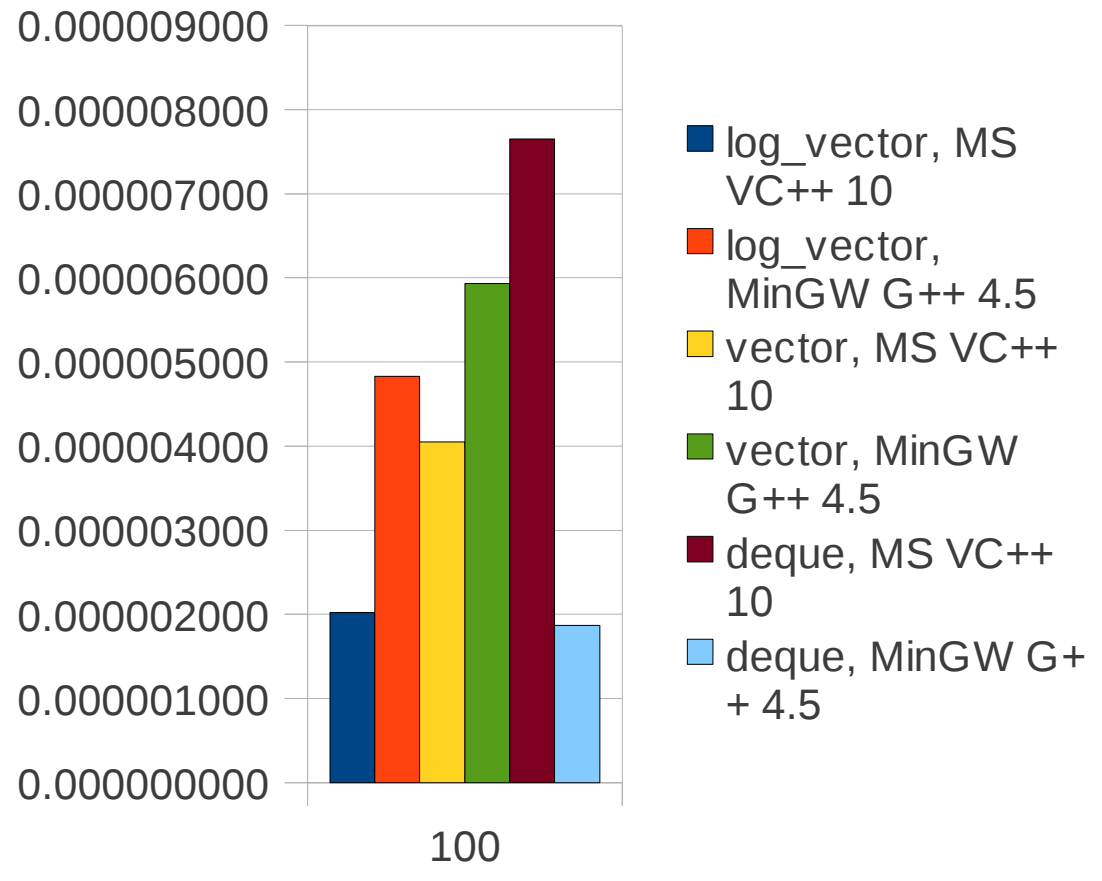

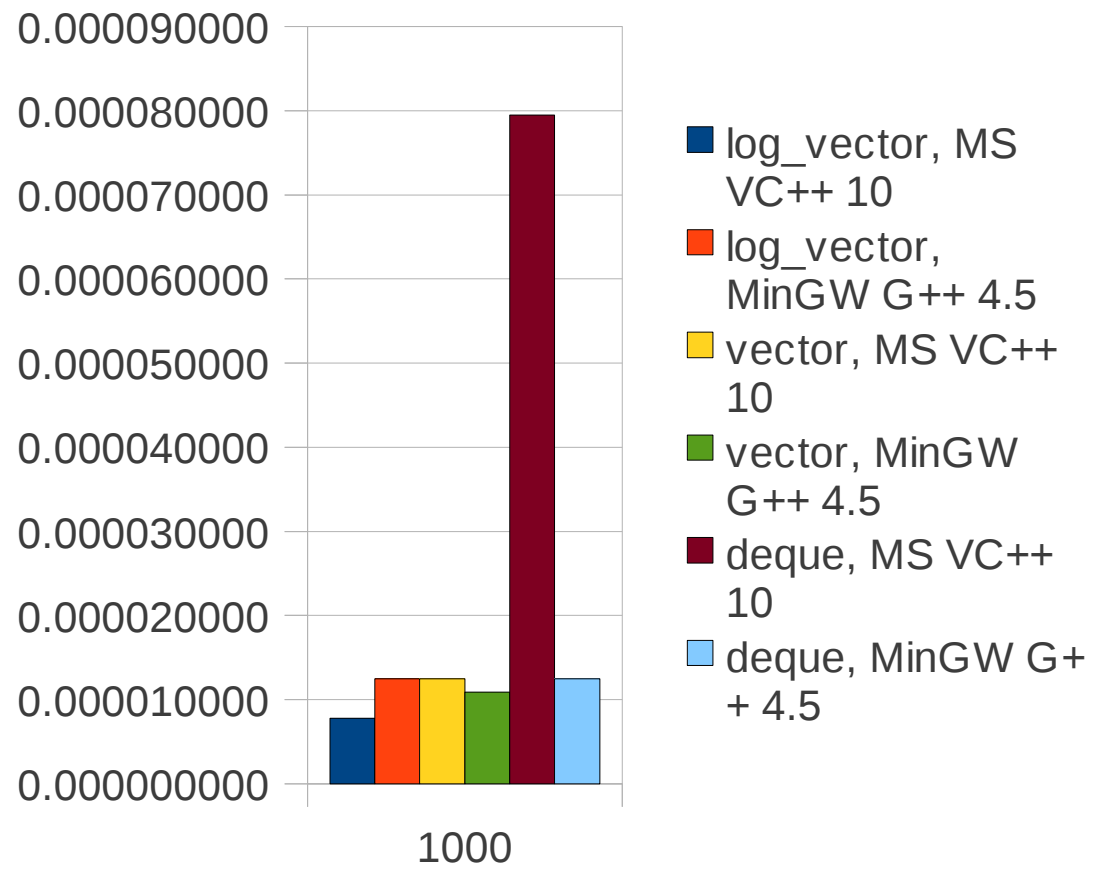

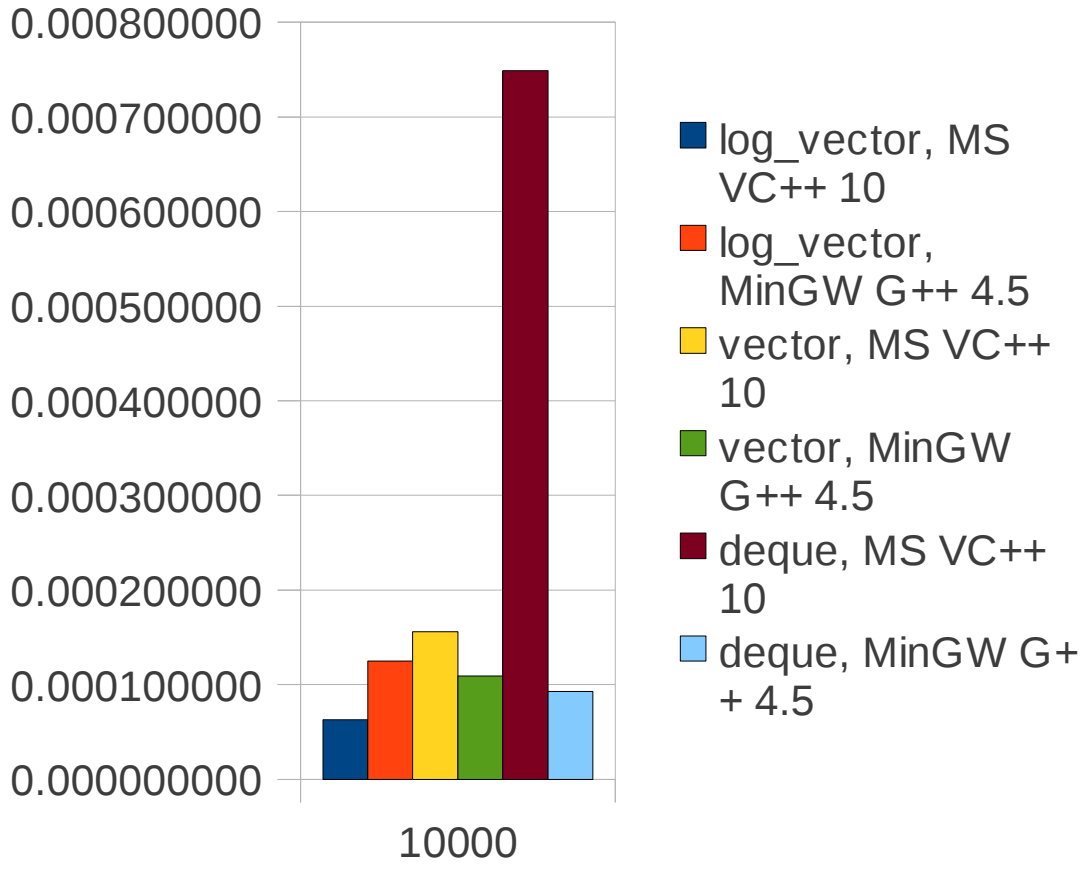

Only two rival data structures with elements of a large size were tested – log_vector and deque's GNU version. Results are the following.

| N | log_vector, MS V | deque, MinGW G- |
|---|---|---|
| 100 | 2.020000000 | 1.870000000 |
| 1000 | 0.780000000 | 1.250000000 |
| 10000 | 0.630000000 | 0.930000000 |
| 100000 | 1.250000000 | 1.090000000 |
| 1000000 | 1.400000000 | 1.250000000 |
| 10000000 | 1.400000000 | 1.560000000 |

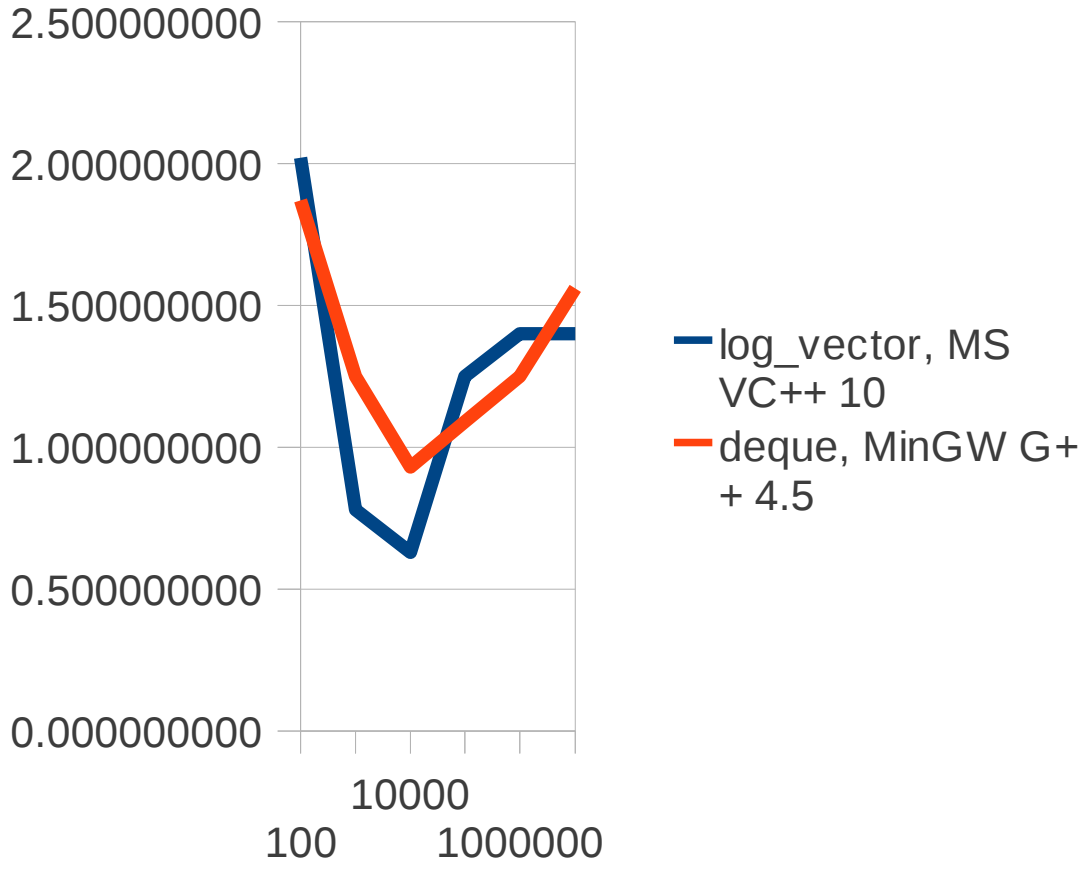

Secondly, the array functionality was tested on the data structures. Results follows.

| N | log_vector, MS V | log_vector, MinG\ | vector, MS VC++ | vector, MinGW G- | deque, MS VC++ | deque, MinGW G- |
|---|---|---|---|---|---|---|
| 100 | 0.000002180 | 0.000004990 | 0.000000460 | 0.000000930 | 0.000010450 | 0.000002180 |
| 1000 | 0.000009400 | 0.000018800 | 0.000003200 | 0.000003100 | 0.000104600 | 0.000015600 |
| 10000 | 0.000093000 | 0.000187000 | 0.000015000 | 0.000032000 | 0.001014000 | 0.000125000 |
| 100000 | 0.001410000 | 0.001250000 | 0.000310000 | 0.000310000 | 0.010600000 | 0.001400000 |
| 1000000 | 0.017200000 | 0.020300000 | 0.011000000 | 0.010900000 | 0.128000000 | 0.023400000 |
| 10000000 | 0.234000000 | 0.219000000 | 0.109000000 | 0.094000000 | 1.326000000 | 0.250000000 |

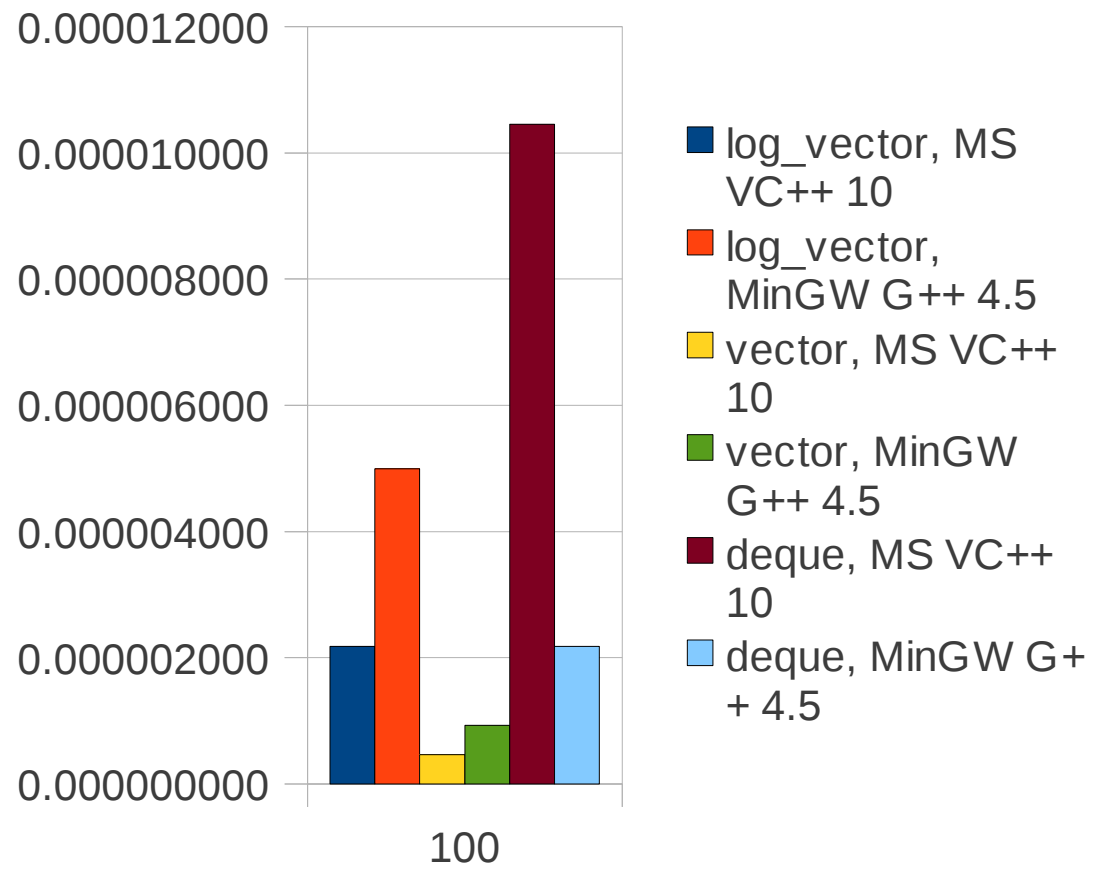

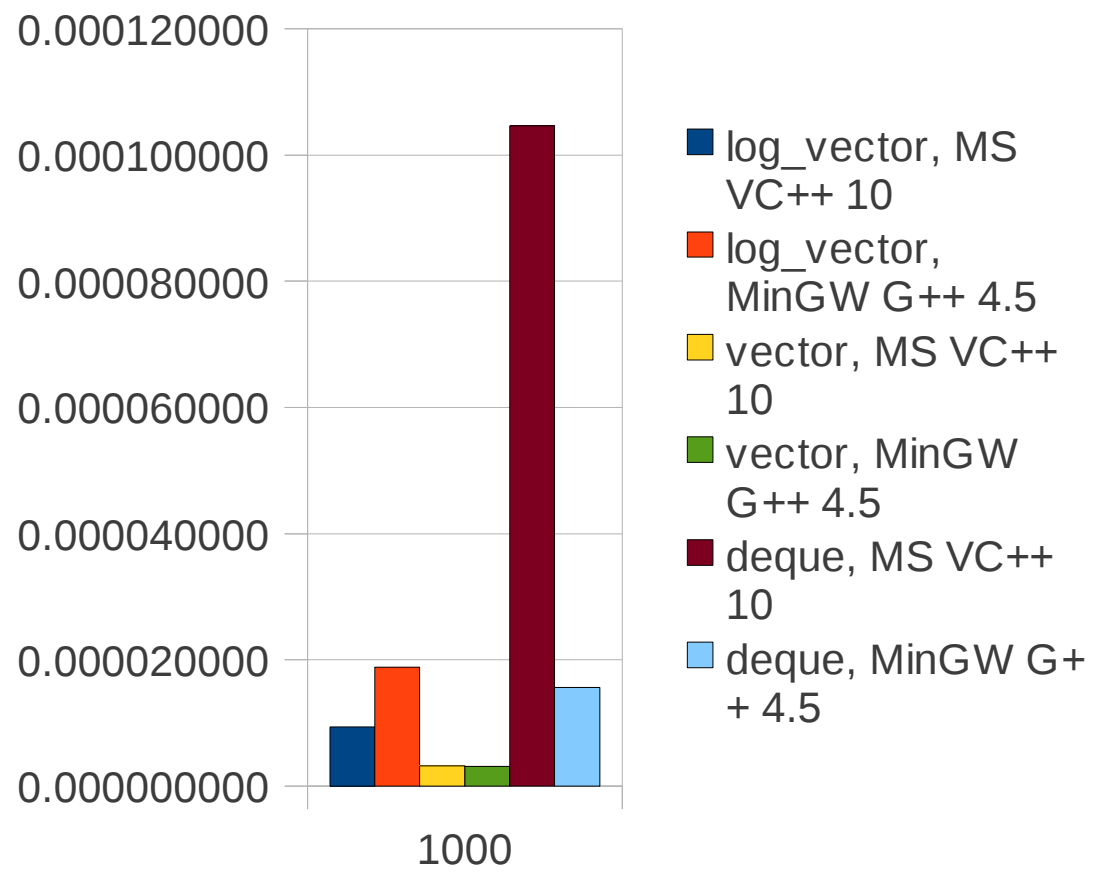

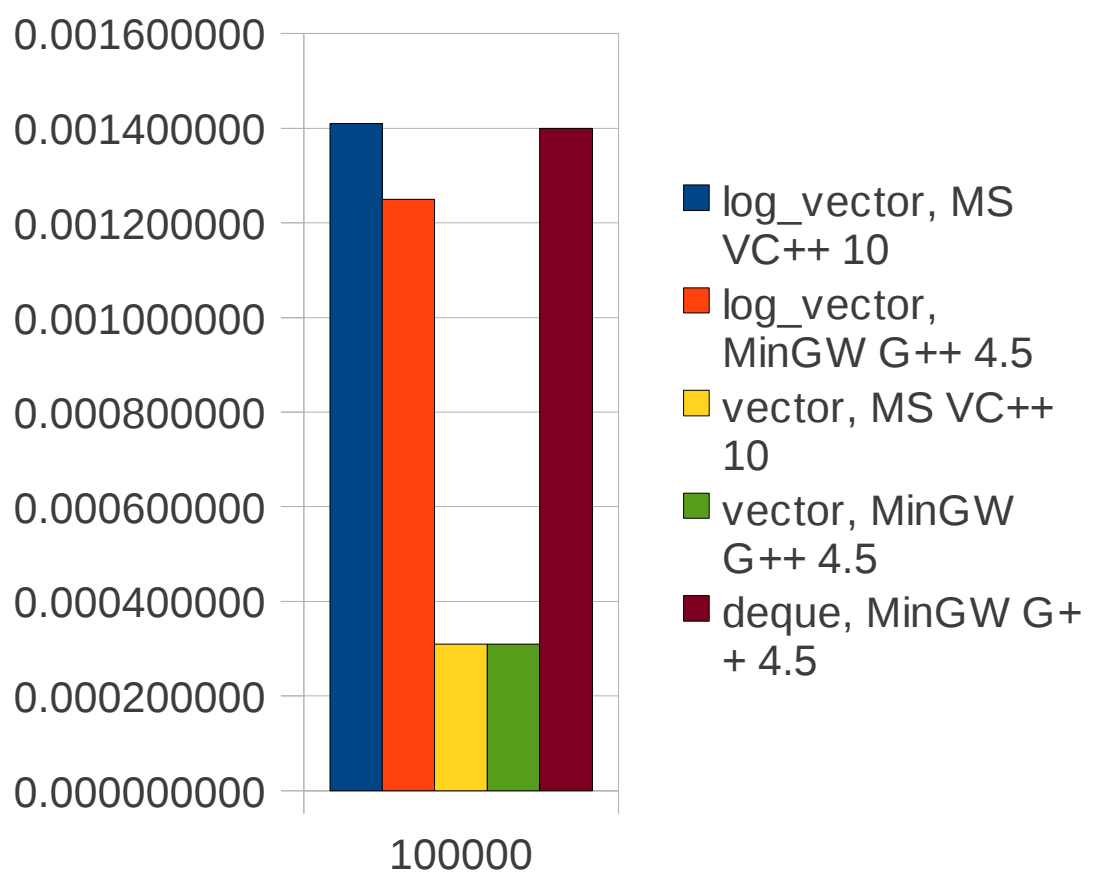

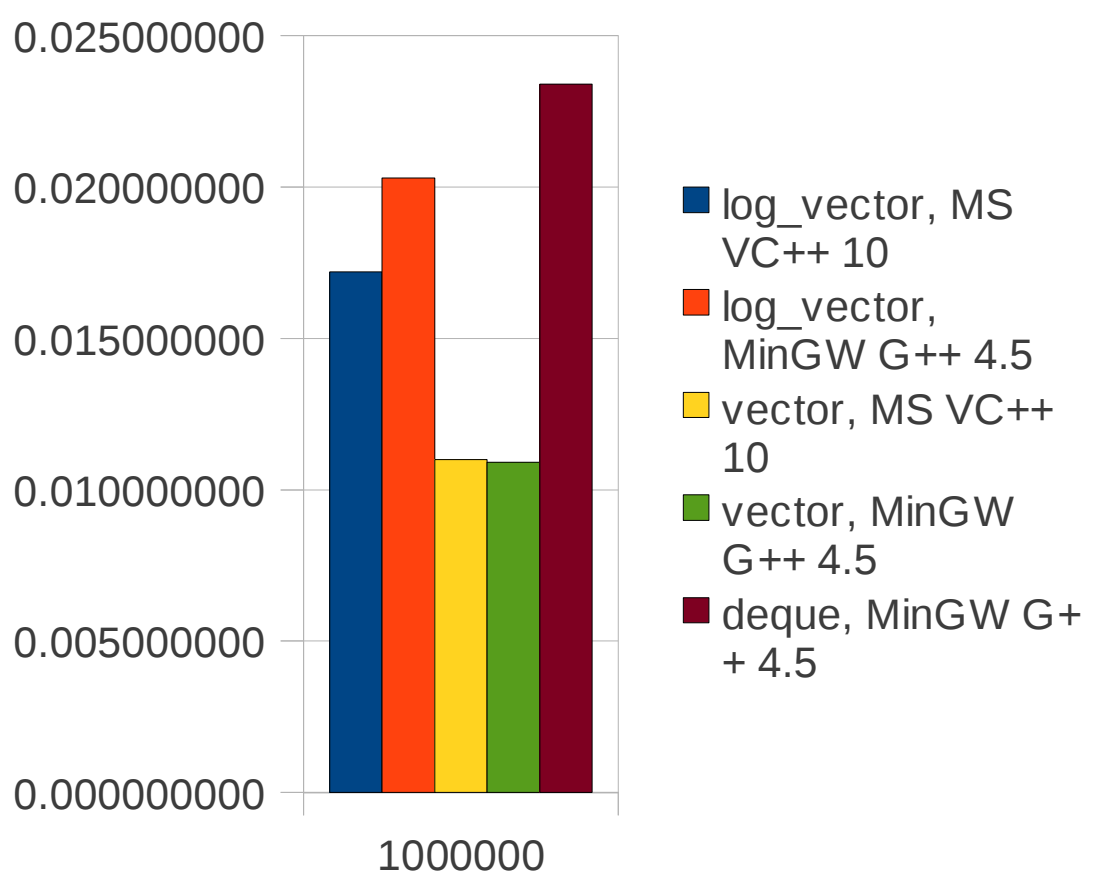

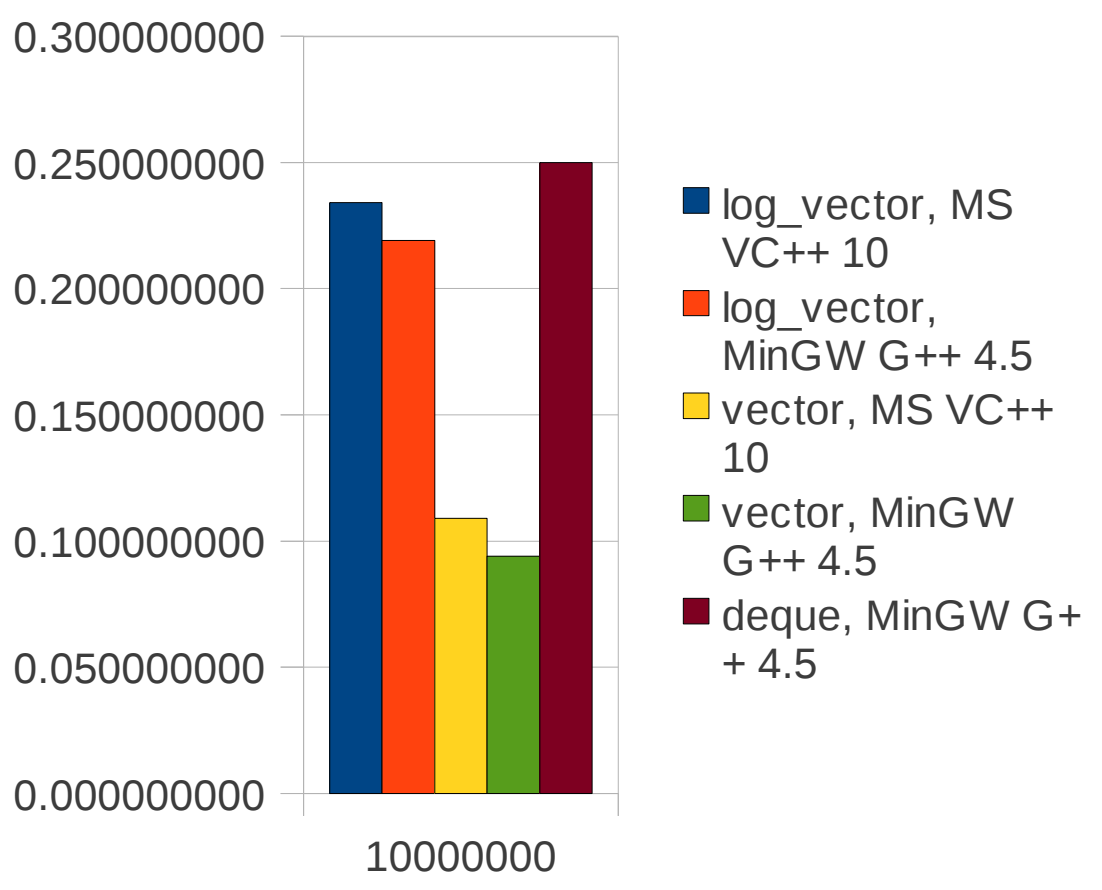

The same tests were run on the large data structure element's sizes. Only two rivals – log_vector and GNU deque were compared. Results follow.

| N | log_vector, MS V | deque, MinGW G- |
|---|---|---|
| 100 | 2.180000000 | 2.180000000 |
| 1000 | 0.940000000 | 1.560000000 |
| 10000 | 0.930000000 | 1.250000000 |
| 100000 | 1.410000000 | 1.400000000 |
| 1000000 | 1.720000000 | 2.340000000 |
| 10000000 | 2.340000000 | 2.500000000 |

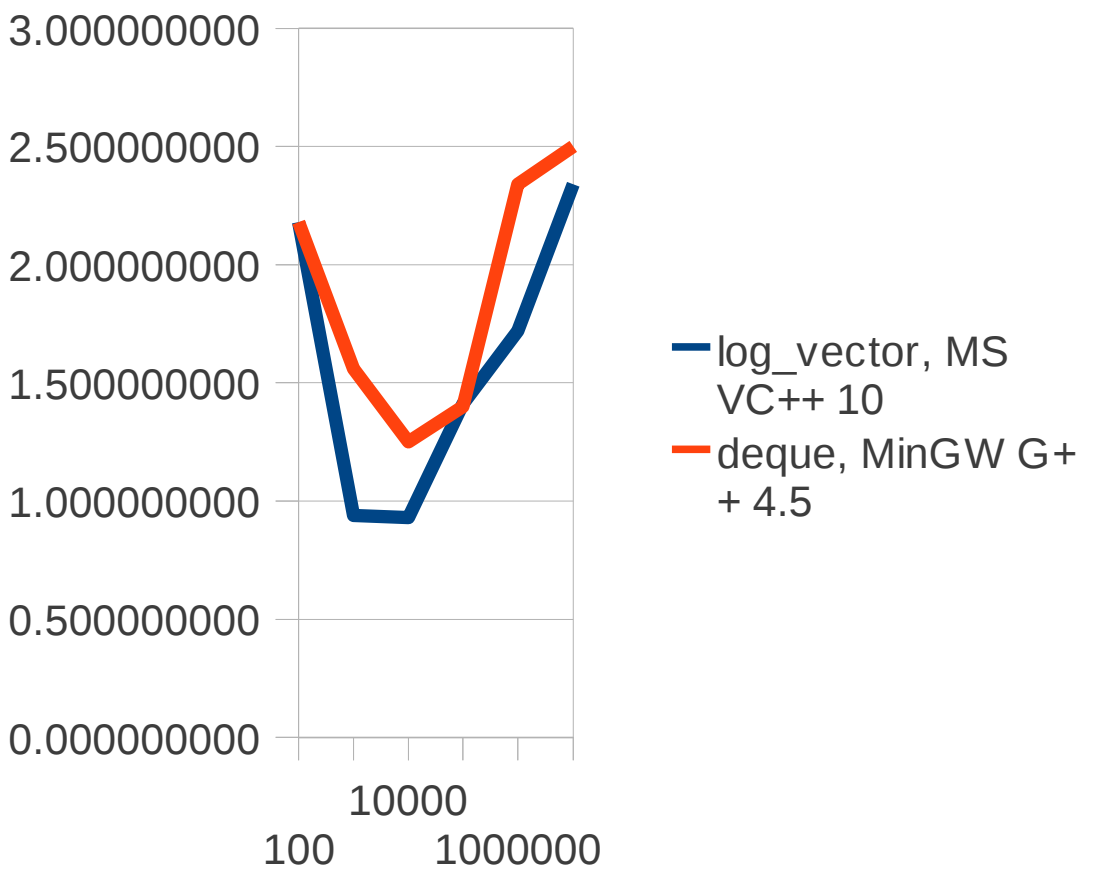

Data structure creation time was also analyzed. Results follows.

| N | log_vector, MS V | log_vector, MinG\ | vector, MS VC++ | vector, MinGW G- | deque, MS VC++ | deque, MinGW G- |
|---|---|---|---|---|---|---|
| 100 | 0.000001710 | 0.000004210 | 0.000000310 | 0.000000620 | 0.000007330 | 0.000001560 |
| 1000 | 0.000003100 | 0.000006200 | 0.000003100 | 0.000001600 | 0.000074900 | 0.000006200 |
| 10000 | 0.000016000 | 0.000031000 | 0.000016000 | 0.000015000 | 0.000686000 | 0.000031000 |
| 100000 | 0.000150000 | 0.000630000 | 0.000160000 | 0.000150000 | 0.007020000 | 0.000310000 |
| 1000000 | 0.004700000 | 0.006200000 | 0.004700000 | 0.006300000 | 0.081100000 | 0.007800000 |
| 10000000 | 0.078000000 | 0.078000000 | 0.047000000 | 0.047000000 | 0.843000000 | 0.078000000 |

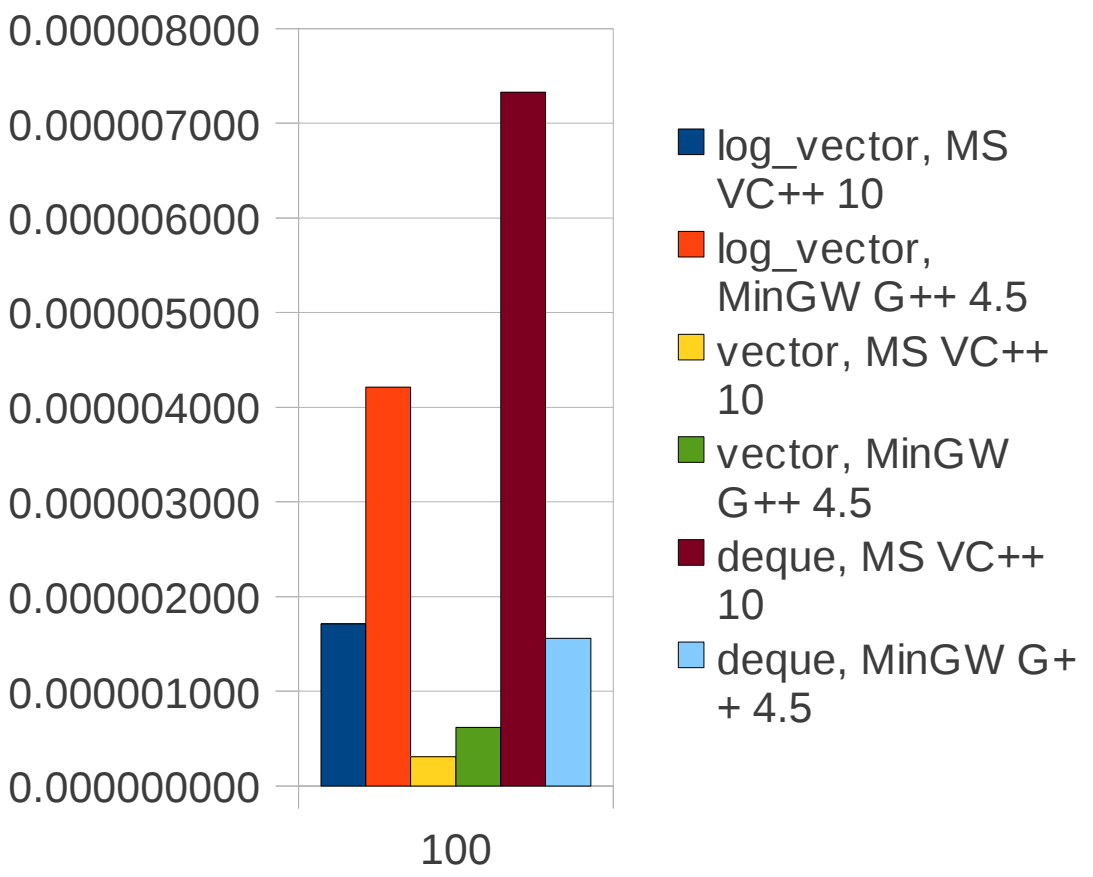

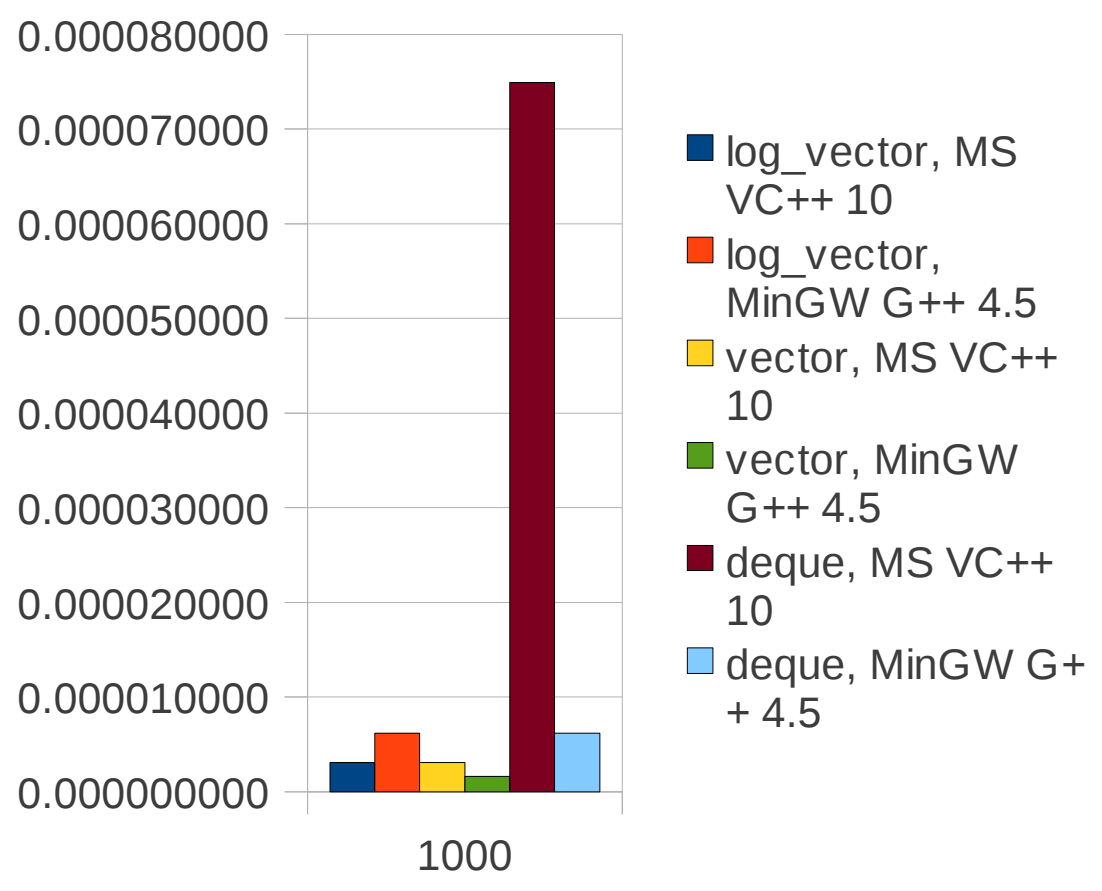

MS deque implementation was removed from the tests, because it was a clear outsider on the hardware used.

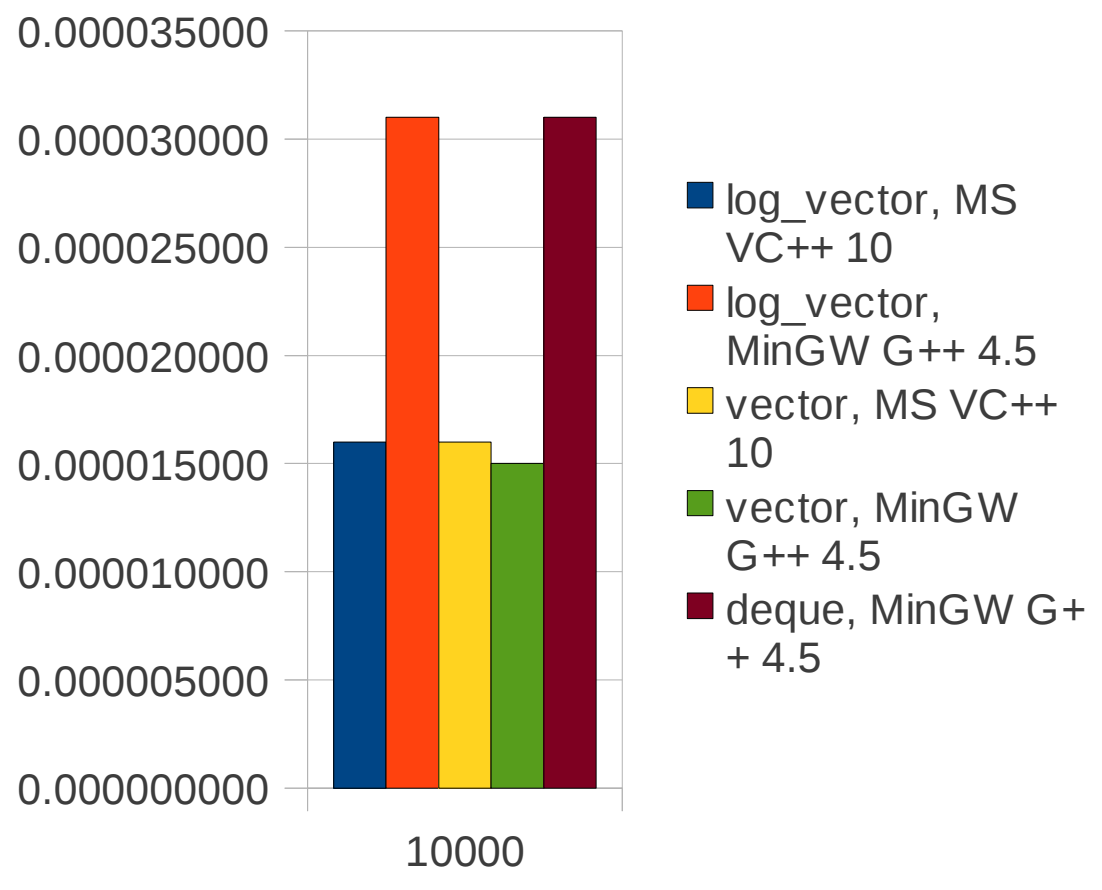

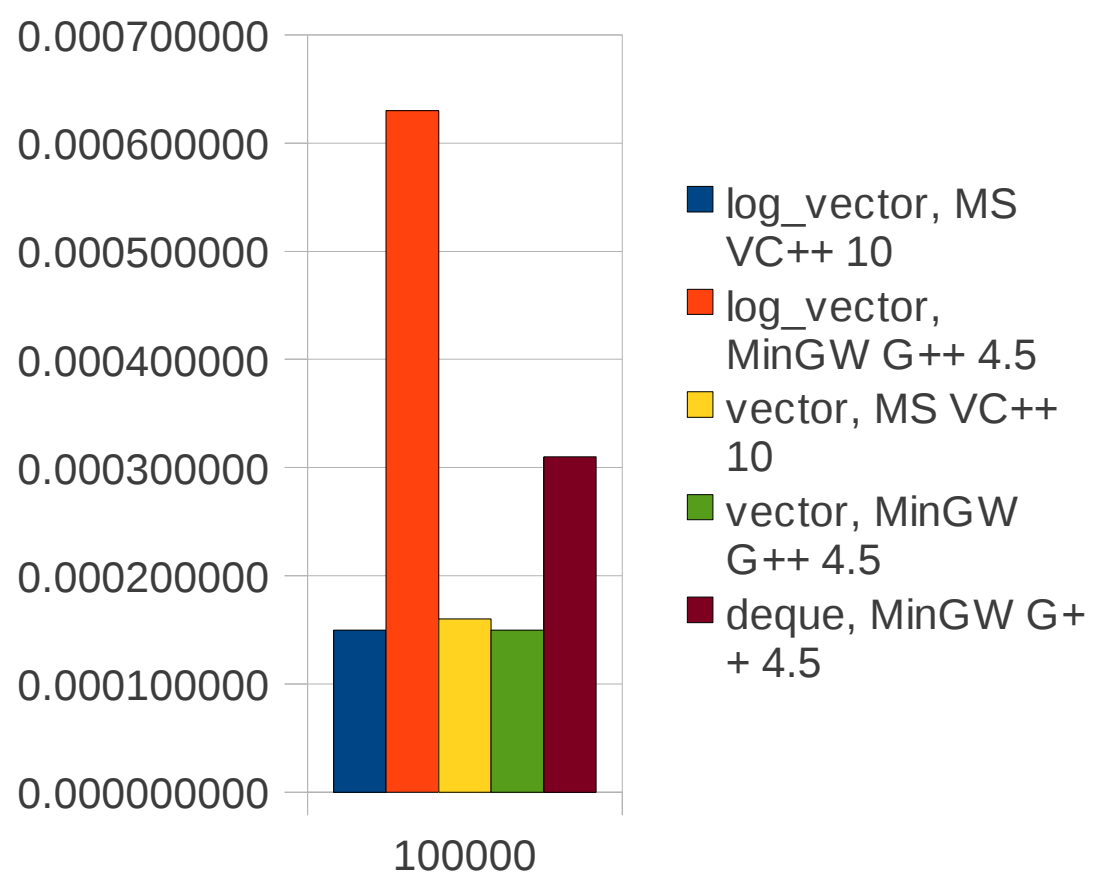
0.000700000
0.000600000
0.000500000
0.000400000
0.000300000
0.000200000
0.000100000
0.000000000
100000
log_vector, MS VC++ 10
log_vector, MinGW G++ 4.5
vector, MS VC++ 10
vector, MinGW G++ 4.5
deque, MinGW G++ 4.5

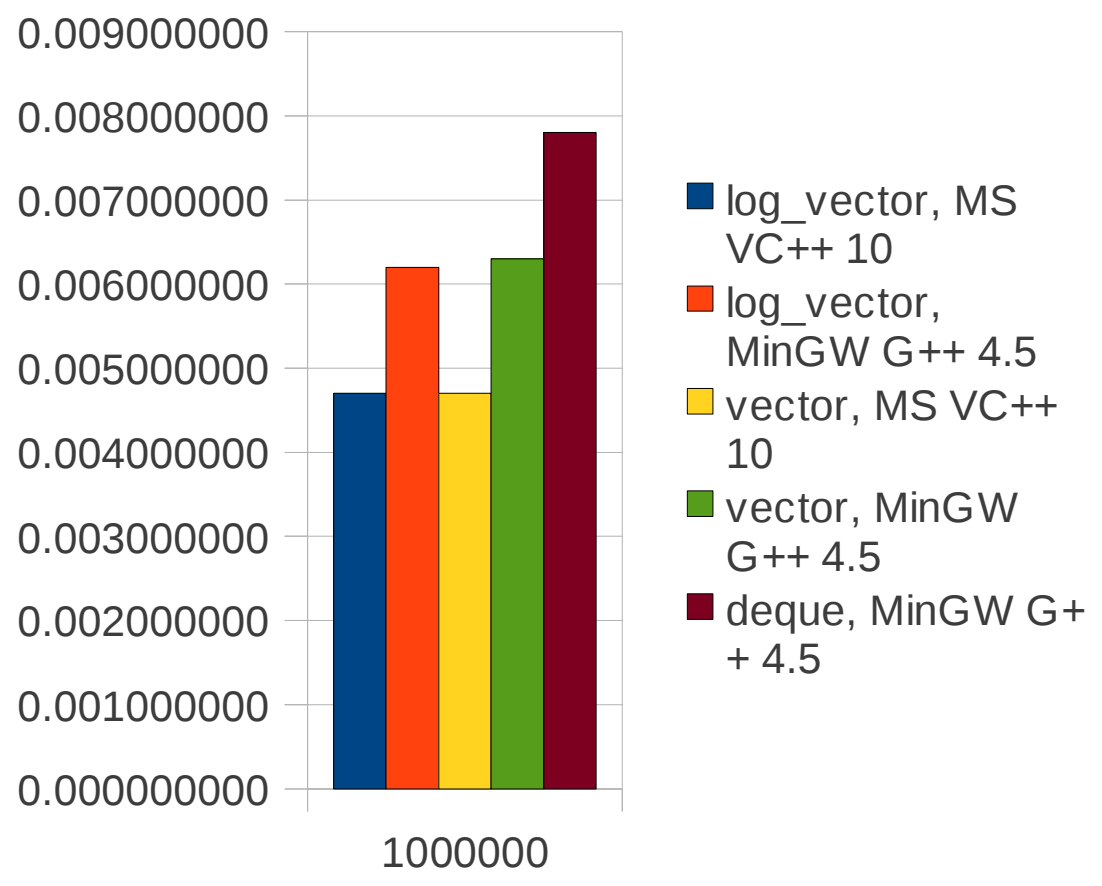
0.009000000
0.008000000
0.007000000
0.006000000
0.005000000
0.004000000
0.003000000
0.002000000
0.001000000
0.000000000
1000000
log_vector, MS VC++ 10
log_vector, MinGW G++ 4.5
vector, MS VC++ 10
vector, MinGW G++ 4.5
deque, MinGW G++ 4.5

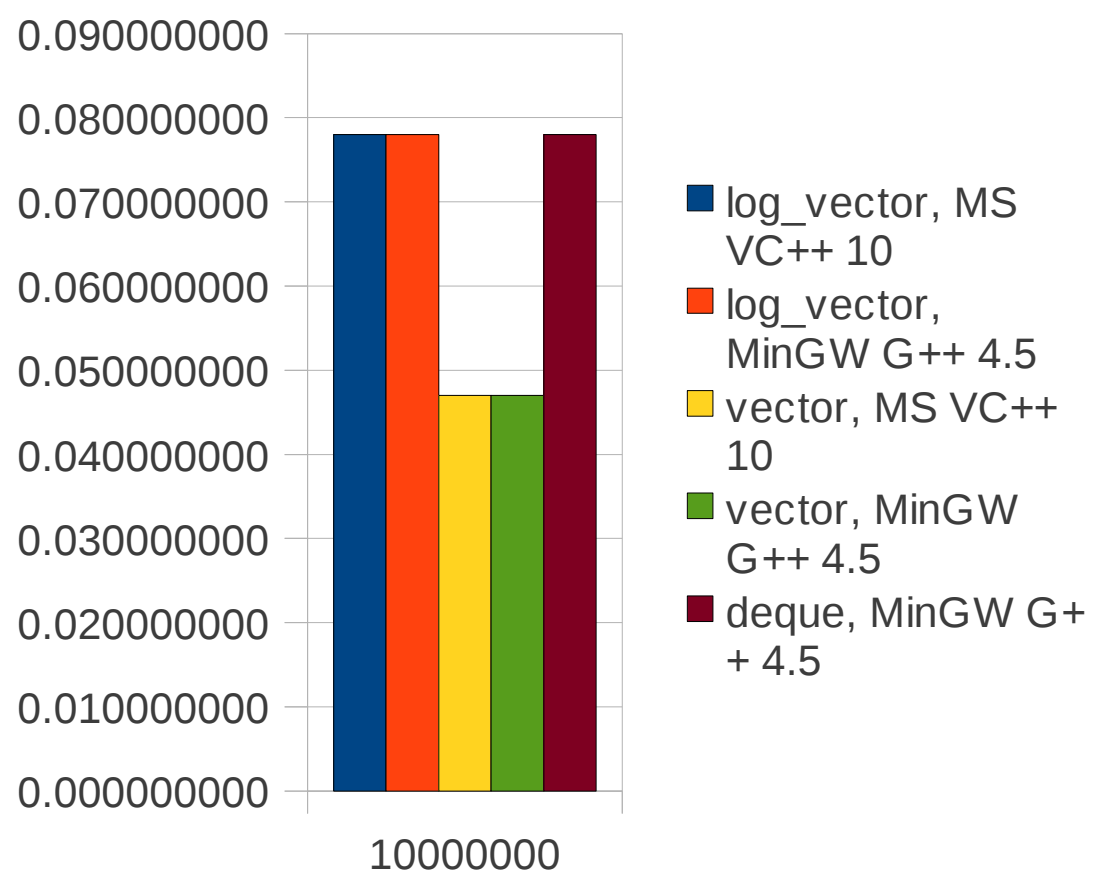
0.090000000
0.080000000
0.070000000
0.060000000
0.050000000
0.040000000
0.030000000
0.020000000
0.010000000
0.000000000
10000000
log_vector, MS VC++ 10
log_vector, MinGW G++ 4.5
vector, MS VC++ 10
vector, MinGW G++ 4.5
deque, MinGW G++ 4.5

## Conclusion

The described data structure can be suggested when both stack-like functionality and element indexing are necessary, or when data type size is large, instead of the deque.

Nevertheless, in order to be a rival for the deque, it is necessary to optimize the creation of the small log_vector-s. The idea is ti store the first 3-4 levels in a one memory block, and to store shifted references to this memory block.